\documentclass[fleqn,10pt]{wlscirep}
\usepackage[utf8]{inputenc}
\usepackage[T1]{fontenc}
\usepackage{url}
\usepackage{lineno}
\usepackage{subfig}
\usepackage{multirow}
\usepackage{multicol}
\usepackage{comment}
\usepackage{amsmath}
\usepackage[T1]{fontenc} 
\usepackage{xcolor}
\newcommand{\pT}{${\rm p_T}$}

\newcommand{\deta}{$\Delta \eta$}
\newcommand{\dphi}{$\Delta \phi$}

\usepackage{hyperref}
\usepackage{lineno}
\usepackage{csquotes}

\title{Trees versus Neural Networks for enhancing tau lepton real-time selection in proton-proton collisions}

\author[1,2,*]{M. Yaari}
\author[1]{U. Barron}
\author[1,*]{L. Pascual Dom\'inguez}
\author[1]{B. Chen}
\author[1]{L. Barak}
\author[1]{E. Etzion}
\author[2]{R. Giryes}

\affil[1]{School of Physics and Astronomy, Tel Aviv university, Ramat Aviv, 69978, Israel}
\affil[2]{School of Electrical Engineering, Tel Aviv university, Ramat Aviv, 69978, Israel}

\affil[*]{yaari.maayan@gmail.com, luis.pascual@cern.ch}

\begin{abstract}
 This paper introduces supervised learning techniques for real-time selection (triggering) of hadronically decaying tau leptons in proton-proton colliders. By implementing classic machine learning decision trees and advanced deep learning models, such as Multi-Layer Perceptron or residual neural networks, visible improvements in performance compared to standard threshold tau triggers are observed. We show how such an implementation may lower 
 selection energy thresholds, thus contributing to increasing the sensitivity of searches for new phenomena in proton-proton collisions classified by low-energy tau leptons. Moreover, we analyze when it is better to use neural networks versus decision trees for tau triggers with conclusions 
 relevant to other problems in physics.
\end{abstract}

\begin{document}

\flushbottom
\maketitle
%
%
\thispagestyle{empty}

\section*{Introduction}
\label{intro}

High-energy physics experiments currently face the challenge of recording massive volumes of data at an extremely high rate. 
The Large Hadron Collider (LHC), a particle accelerator that collides proton bunches at a 40 MHz frequency, exemplifies this issue. 
The energy deposits  and trajectories of particles from proton-proton ($pp$) collisions  are recorded by experimental apparatuses, such as the ATLAS and CMS experiments~\cite{ATLAS:2008xda,CMS:2008xjf}.
These processes result in a data load that is difficult to manage within stringent timing constraints. 
 In addition to these technical challenges,
while the majority of events contain well-understood physical phenomena that are less critical to study, a minuscule portion of the ($pp$) collisions unveil rare physics essential for storage and in-depth analysis. 
Therefore, given the importance of not overlooking these rare events, an extremely efficient filtering mechanism, known as the "trigger system," is employed.  
 The trigger system promptly decides, with limited real-time information, which events warrant storage and further analysis, whereas the remainder are discarded ~\cite{ATLAS:2016wtr}.

The trigger system for ATLAS and CMS consists of multiple stages. The first level (L1) of a multi-level trigger system, which is generally present in hadron collider experiments~\cite{ATLAS-CONF-2017-061,CMS:2016ngn}, uses stringent constraints to select the events that their data are kept  for the following trigger steps. 
To speed up the process, the L1 trigger is ordinarily implemented on dedicated electronic circuits, while recent implementations use field-programmable gate arrays (FPGAs), allowing a given algorithm to respond promptly.
Whether or not an event passes the trigger depends on signals recorded in the different detector subsystems.

Several trigger paths (chains) run in parallel, each with a predefined rate and bandwidth. A key feature in many of the chains is the energy deposited in the calorimeter. Until recently, simple threshold-base algorithms have been utilized to detect significant energy deposits within the calorimeters, supplemented by isolation criteria to mitigate background rates. Events passing the L1 trigger are further filtered by the high-level trigger (HLT), based on CPUs and GPUs, where more advanced algorithms can be used.
However, there have been recent developments in engineering that allow for the implementation of machine learning (ML) methods on FPGAs within the timing and resource constraints necessary at L1~\cite{Duarte:2018ite, Loncar:2020hqp, Carlson:2022dgb}. These advancements represent a paradigm shift, empowering L1 with the capability to execute advanced algorithms previously exclusive to the HLT, thereby enhancing event selection efficiency and broadening the scope of physics analyses within the experiments.

\begin{figure}[ht]
\centering  
\includegraphics[width=0.30\columnwidth]{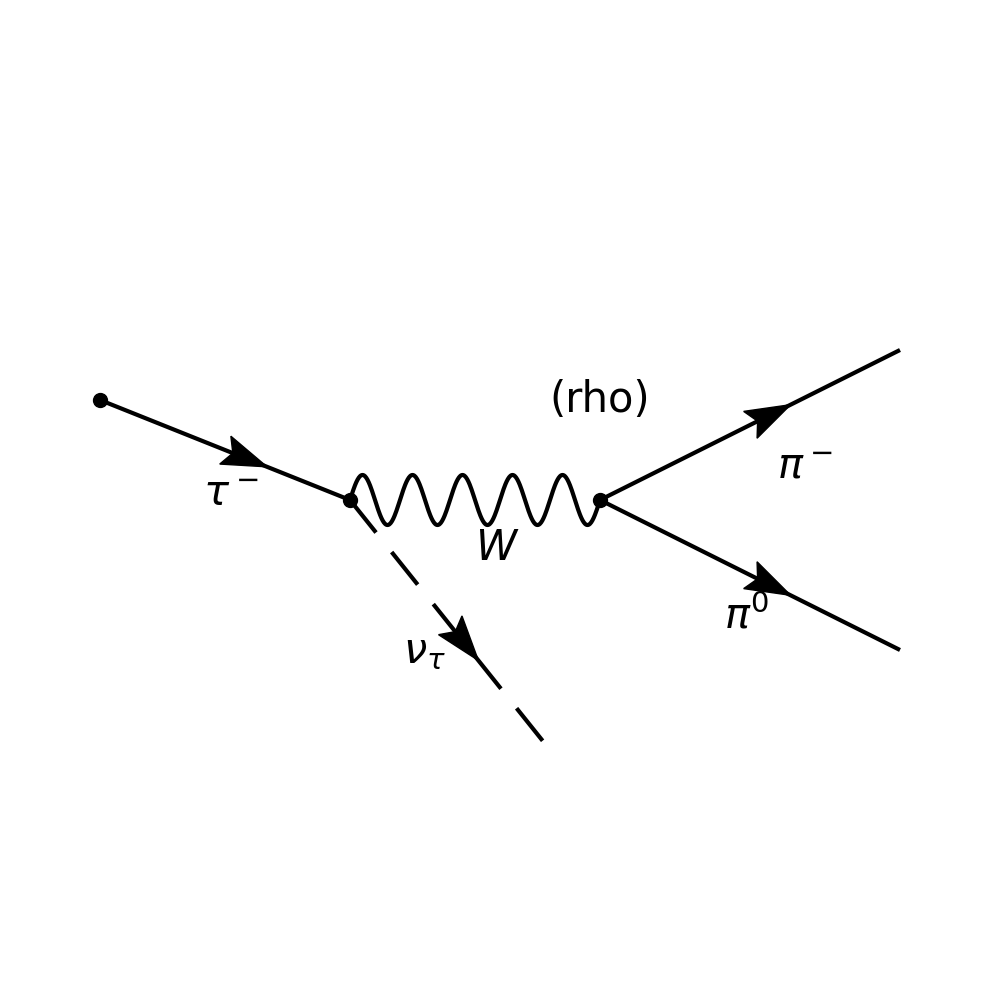}
\includegraphics[width=0.30\columnwidth]{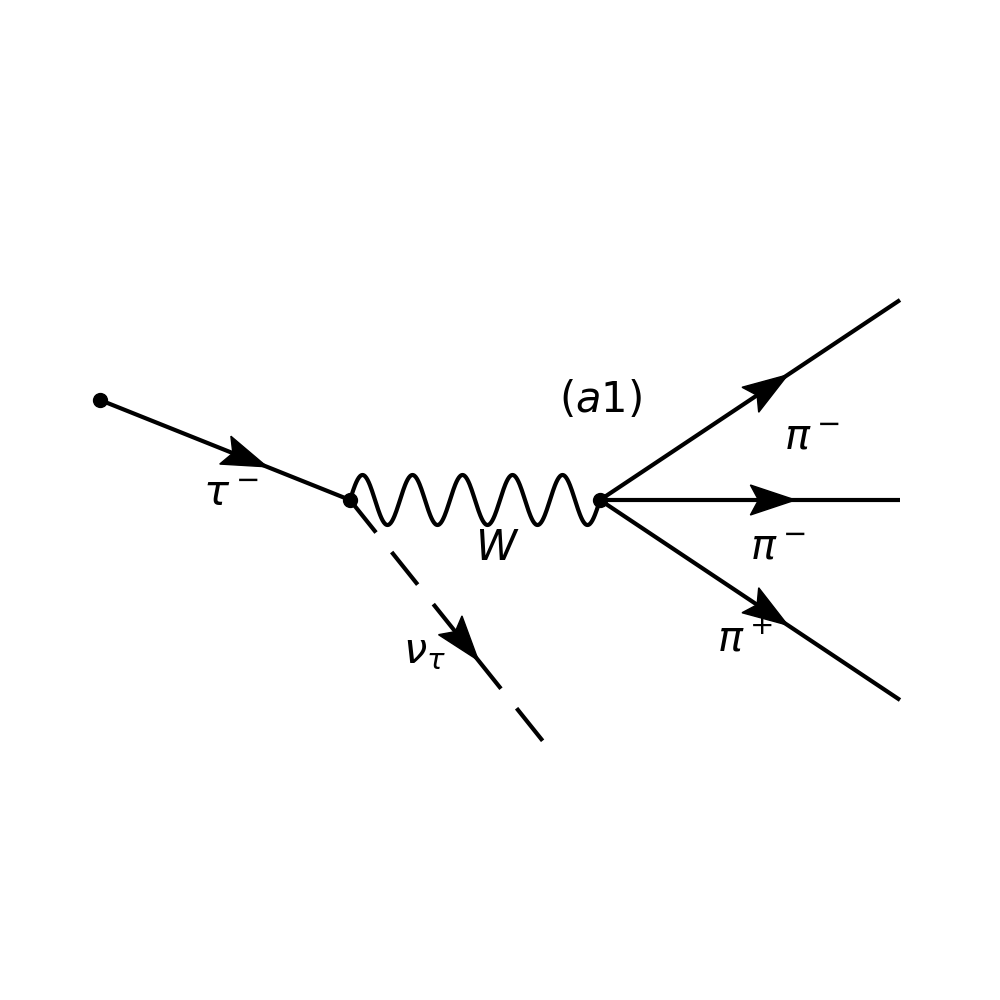}
\includegraphics[width=0.30\columnwidth]{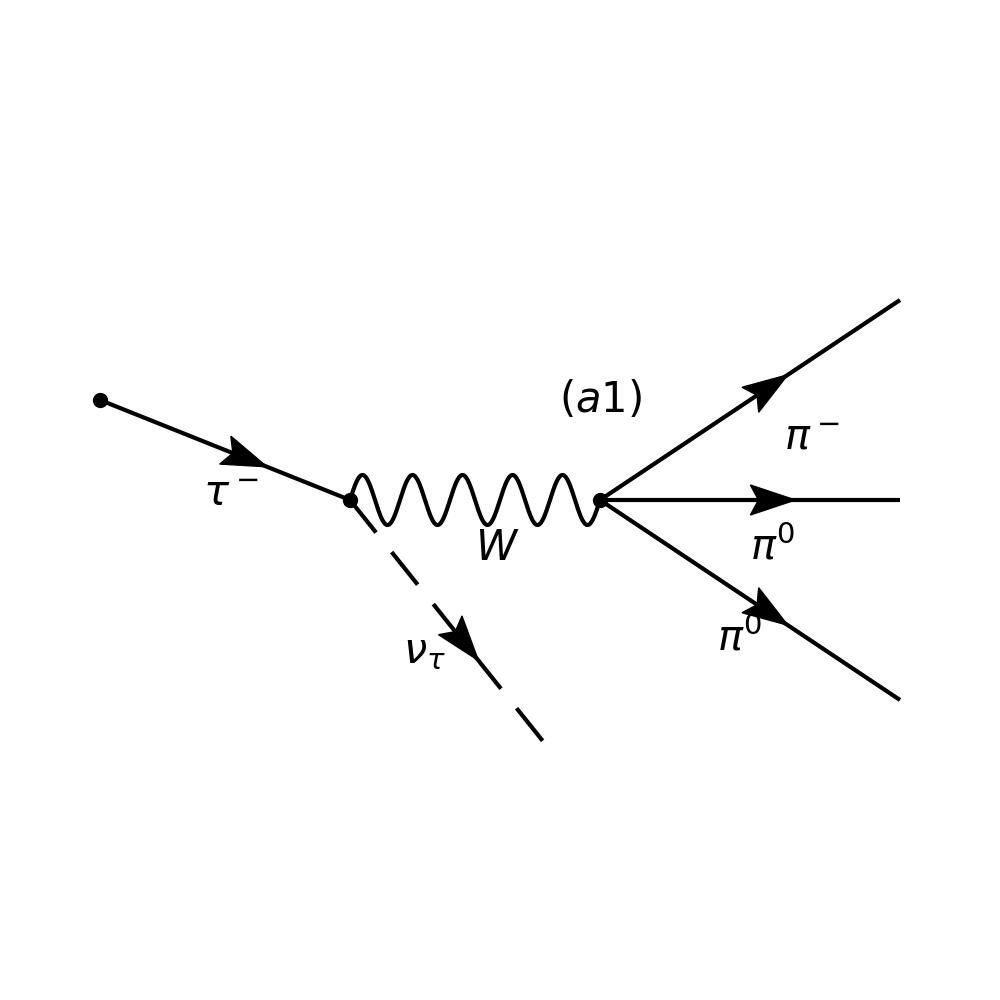}
  
   \caption{Feynman diagrams of the most frequent hadronic decays of the $\tau$ lepton. The probability of a $\tau$ to decay into hadronic states is 64.79\%.  Out of which (from left to right) 40\% are $\tau \rightarrow \pi^-\pi^0$ decays, 14.37\% decay into $\pi^-\pi^-\pi^+$ and 14.29\% are $\tau\rightarrow \pi^-2\pi^0$ decays
~\cite{ParticleDataGroup:2022pth}.
  }
  \label{fig:feynman}
\end{figure}

When searching for new physics phenomena, theoretically motivated processes that fall within the aforementioned category include, for instance, those involving hypothesized light particles from so-called hidden sectors. These particles may have couplings to the Standard Model (SM) fields that are proportional to their mass~\cite{Branco:2011iw,Djouadi:2005gj}. 
Consequently, $pp$  collision events carrying third generation fermions such as the b-quarks or the tau ($\tau$) leptons   are attractive to analyze  since they could be sensitive to physics Beyond the Standard Model (BSM) scenarios.  
However, despite the strong theoretical motivation for storing and analyzing events involving tau leptons, distinguishing between them and the majority of hadronic jet events in a $pp$ collider poses a significant challenge. This difficulty primarily arises from the tendency of the majority of the tau leptons to decay into hadronic final states (see Fig.~\ref{fig:feynman}).

The similarity between the tau hadronic decays and the overwhelming background from jets produced in the hadronization of quarks and gluons characterizing $pp$ collisions significantly complicates the selection for tau events (Fig~\ref{fig:tau-had-jets}).
This challenge is notably acute in the low-energy regime, where the signatures of the two phenomena are almost indistinguishable.   
Although identification algorithms that leverage advanced deep learning techniques have been developed to distinguish between these particular signatures, as reported in~\cite{CMS:2022prd}, they are not suitable for application during the initial stages of the trigger process. This is due to the stringent latency constraints and the limited information that is available at these early stages.

The adoption of FPGA technology in upgrading  the tau trigger 
promises to enhance algorithmic complexity and effectiveness beyond the capabilities of currently used methods.

Within this paper, we explore various machine learning algorithms, highlighting the merits and limitations of each. 
The ATLAS collaboration has scheduled the tau trigger system upgrade in two phases—Phase 1 and Phase 2. Our discussion sets the stage for potential breakthroughs expected with the Phase 1 upgrade.
A notable enhancement in the tau trigger system during the Phase 1 upgrade is the introduction of finer detector granularity. To mimic ATLAS's data environment, we generate synthetic data that emulates different levels of detector granularity. This enables us to test basic machine learning algorithms on varied granular structures to determine how detector granularity affects algorithm choice. The uncertainty regarding the definitive selection of hardware for the trigger system leads us to evaluate the real-time applicability of each algorithm. We conduct this assessment through a comparative analysis focused on algorithmic parameters indicative of memory consumption. The presented algorithms are tailored for 3D data outputs from the experiment's calorimeter system, inspired by computer vision techniques applicable to both 2D and 3D image analysis tasks.  



This paper is organized as follows. It starts with providing a comprehensive overview of the experimental aspects involved in the triggering of tau leptons in $pp$  colliders, 
setting the stage for the introduction of 
the dataset used in the NN training. Than it presents the performance outcomes of the various NN architectures studied.  
Finally, it is concluded  by summarising the key findings and discussing future directions 
 for implementing NN algorithms in fast data processing. 
 The complete implementation of this work can be accessed at~\cite{code}.
\section*{Samples and experimental context} 
This section established the experimental framework essential for comprehending the findings presented in this paper. 
It comprises two main parts: firstly, an explanation of the detection process for hadronically decaying tau leptons within the ATLAS experiment, including a presentation of the dedicated ATLAS L1 trigger designed specifically for these events. 
Secondly, a comprehensive description of the signal and background samples utilized in our analysis is provided, offering detailed insights into the experimental methodology strengthening the foundation of this work.

\subsection*{Tau leptons in ATLAS}

Hadronically decaying tau leptons correspond to 64.79\% of its decay branching fraction~\cite{ParticleDataGroup:2022pth}, making this channel critical to many interesting processes such as  the $H\rightarrow \tau\tau$ channel. 
These decays occur through the weak interaction resulting in a tau neutrino, which remains undetected due to its minimal interaction, and hadrons, which are almost fully contained in the calorimetric system~\cite{Achenbach:2008zzb}. 
The calorimeter system is located around\footnote{ATLAS adopts a right-handed coordinate system centered at the interaction point within the detector. The x-axis extends towards the LHC center, the z-axis runs along the beam pipe and the y-axis points upwards orthogonally.  
The polar angle $\theta$ is equivalently expressed as  the pseudorapidity $\eta = - \text{ln } \text{tan} (\theta/2)$. 
 The  x and y axes form the transverse plane, while the azimuthal angle between them is denoted as $\phi$.} the $pp$ collision vertex and it is used to measure the energy deposited by particles while they traverse the material. 
The calorimeter system measures up to $|\eta|$=4.9 and consists of two parts: the electromagnetic calorimeter (ECAL) and the hadronic calorimeter (HCAL). The ECAL is employed to absorb the energy of electromagnetic particles such as electrons and photons, while the HCAL captures the energy of particles involved in hadronic interactions, including hadronic jets, which are the main background in this study.
Each of the calorimeters is longitudinally divided into three layers with different granularities\footnote{An additional layer is present in front of the ECAl to correct for energy losses, denominated presampler or Layer 0.}, to help in the identification and reconstruction of incident particles.

Hadronic jets are the most frequent final state in $pp$ collisions and they represent a challenging background for this work, since no tracking information is available at the L1 trigger, and their energy deposits in the calorimeter mimic the signature of the final state particles produced in a hadronically decaying tau lepton.

The L1 tau trigger in ATLAS has undergone significant evolution over time to adapt to increasingly challenging data-taking conditions and incorporate advancements in firmware development~\cite{ATLAS:2013lic,ATLAS-CONF-2017-061}. Referred to as L1Calo, this trigger relies solely on information obtained from the calorimeter system. Synchronized with $pp$ collisions, L1Calo conducts a scan across all calorimeter channels to pinpoint regions exhibiting substantial energy deposition. 
The progression of the L1Calo trigger system and its algorithms from the data-taking period spanning 2015 to 2018 (Run 2) to the current phase (Run 3) is detailed below.

In order to meet the stringent latency constraints of the L1Calo trigger, calorimeter cells are merged into broader trigger towers spanning both the ECAL and HCAL.
In Run 2, the trigger towers had a granularity of \deta~$\times$\dphi~= $0.1\times 0.1$~\cite{Achenbach:2008zzb} and longitudinally divide per calorimeter. However, in Run 3, a more granular trigger system was available thanks to upgraded trigger processors~\cite{ATLAS:2020osu}, reaching a granularity increase in the $\eta$ direction of up to a factor of 4 in the middle layers of the ECAL~\cite{Aleksa:1602230}.

The search for the most energetic trigger towers is based on constructing 3-dimensional structures known as Trigger Objects (TOBs) within ATLAS calorimeters. Each TOB comprises an $N\times N$ array of trigger towers spanning all layers of the calorimeters, primarily distinguished by its position and energy content. During Run 2, TOBs were characterized by an area of $\Delta\eta \times \Delta\phi = 0.2 \times 0.2$ ($2\times 2$ trigger towers). The selection of TOBs potentially arising from tau candidates relied on a lower transverse energy ($E_{T}$) threshold, calculated as the energy in the transverse plane, coupled with an isolation requirement covering a broader ring, $4\times4$, surrounding the TOB. This isolation condition aimed at discarding candidates with significant energy in the outer ring, characteristic of jets rather than hadronically decaying tau leptons.

Despite enhanced granularity in Run 3, the approach of identifying and selecting tau TOB candidates remained largely similar. Run 3 TOBs now encompass cell regions spanning $\Delta\eta \times \Delta\phi = 0.3 \times 0.3$, with varying granularities in Layers 1 and 2. All trigger towers within this region contribute to the characterization of the signal. Notably, in Run 3, the isolation requirement is separated from the L1Calo step and conducted in a subsequent L1 trigger stage denominated L1Topo. This separation helps manage the number of selected TOBs by L1Calo, maintaining a manageable rate forwarded to L1Topo. Consequently, in the studies discussed here, the objective is to enhance the rejection capabilities before the isolation stage, leveraging the improved granularity of trigger towers to mitigate the rate burden prior to isolation imposition.

\subsection*{Simulated samples}

The task of an L1 tau trigger is to classify $pp$ collision events into those containing hadronically decaying taus (`signal') or those that do not (`background'). 
Events classified as signals are subject to further filtering at the HLT.

Simulated samples are used in this study to mimic the calorimeter structure and replicate the triggering strategy employed by ATLAS on hadronically decaying tau leptons. 
Both signal and background samples are simulated from $pp$ collisions at a center-of-mass energy of 13 TeV with \texttt{MadGraph5\_aMC@NLO} (version 2.9.5)~\cite{Alwall:2014hca} for the matrix element computation and interfaced with Pythia8~\cite{Sjostrand:2014zea} for the decays, hadronization, and underlying event processes. 
The signal sample is composed of events with hadronically decaying tau leptons  obtained from simulated Z boson decays while the background consists of a sample of events with hadronic jet pairs (dijets). 

After the event generation, the smearing of the signals in the calorimeter is performed by Delphes (version 3.4.2)~\cite{Ovyn:2009tx,deFavereau:2013fsa}.
Delphes introduces smearing terms in the detector resolution and other experimental effects, for instance pileup. 
Pileup refers to additional $pp$ collisions simultaneous to the collision of interest. 
The pileup and the interaction between the particles and the material of the calorimeters are modeled with the ATLAS datacard included in the Delphes package~\cite{delphes:github}.

\begin{figure}[ht]
  \centering
  \subfloat[]{
  \includegraphics[width=0.18\columnwidth]{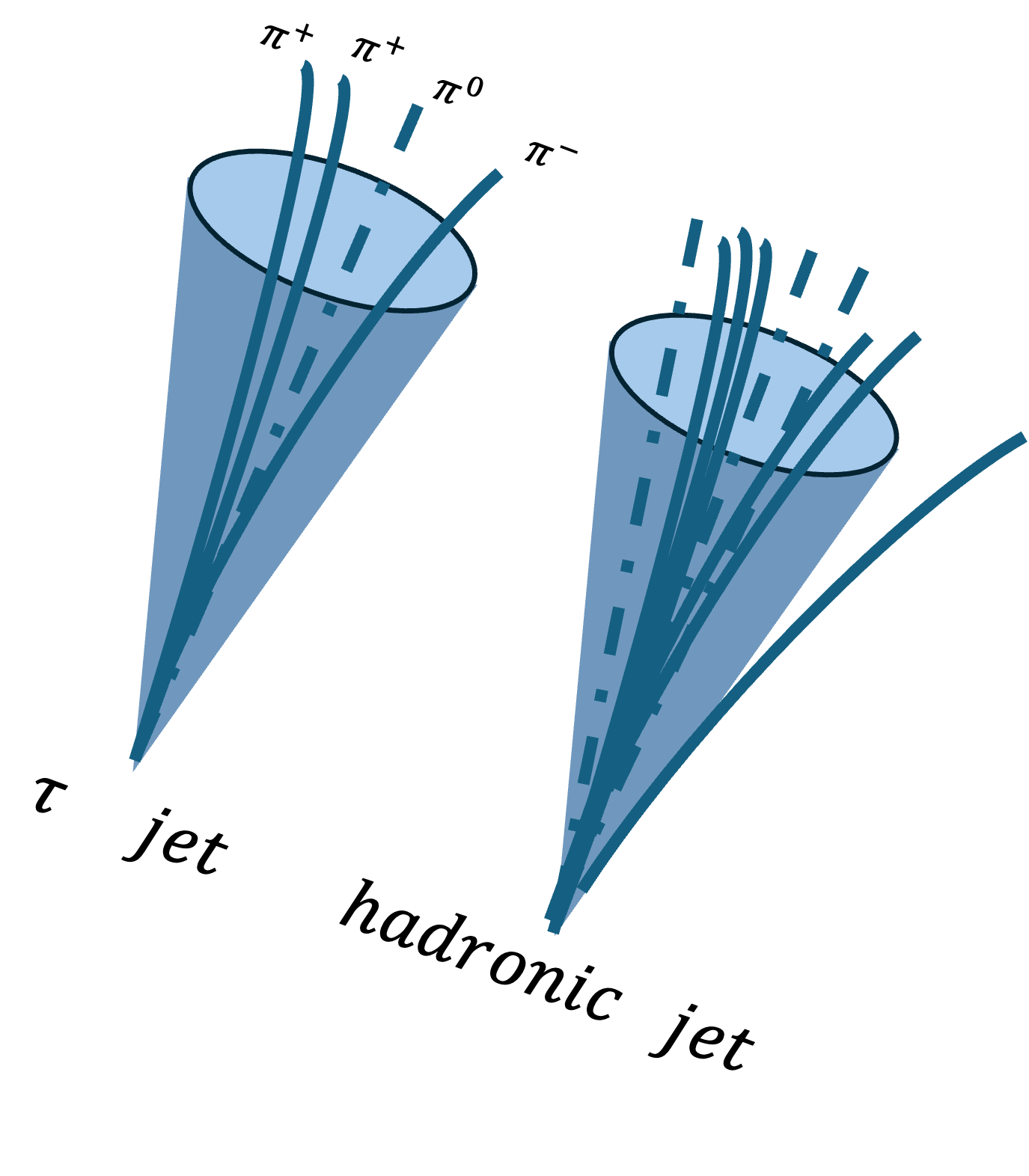}
  \label{fig:tau-had-jets}
  }
  \subfloat[]{
  \includegraphics[width=0.40\columnwidth]{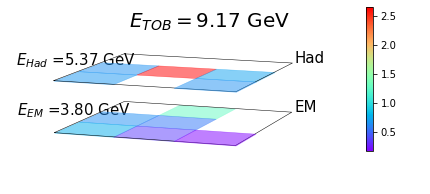}
  \label{fig:tob_visual}
  }
  \subfloat[]{
  \includegraphics[width=0.30\columnwidth]{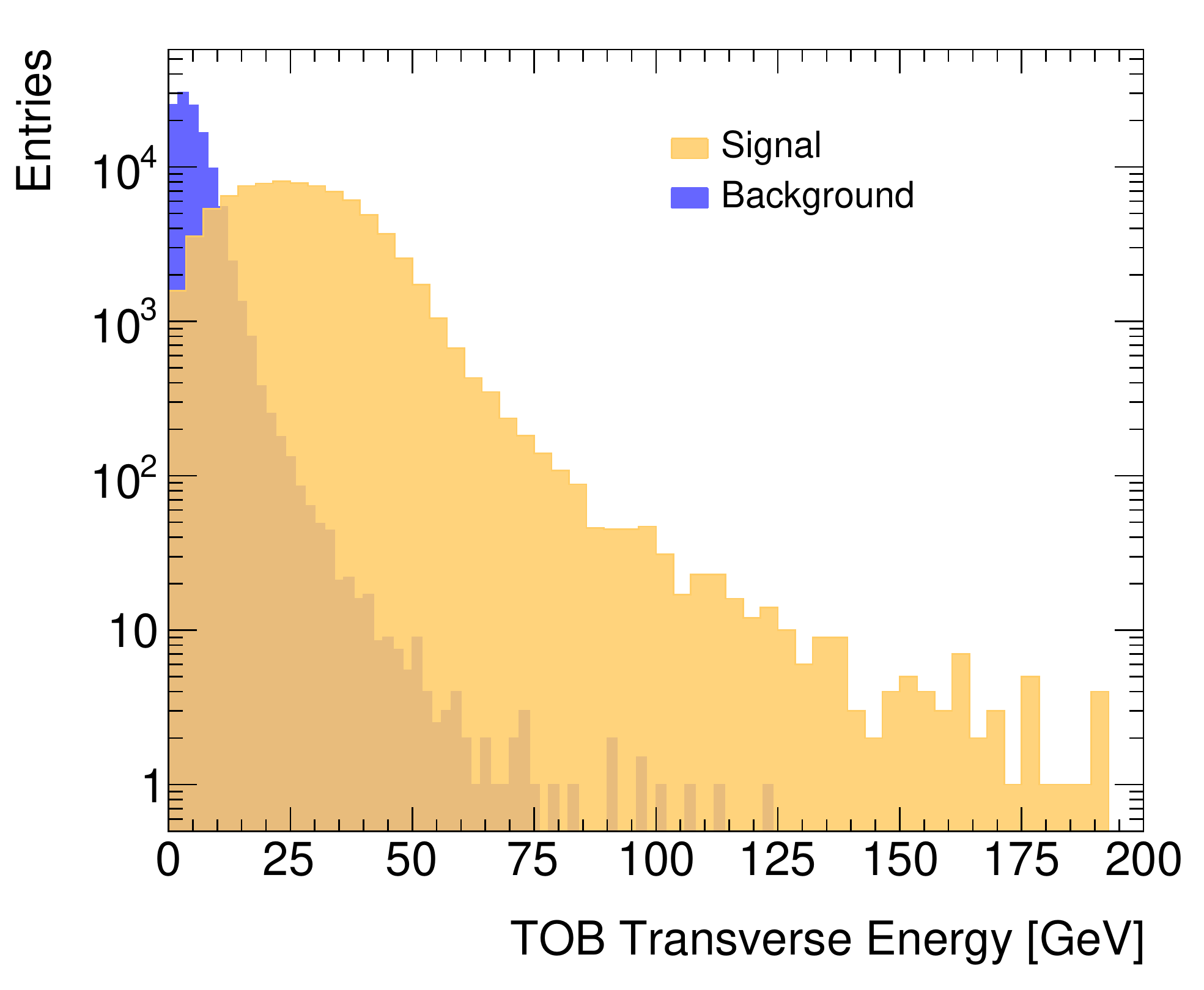}
  \label{fig:pt_tob_comparison}
  }
  \hfill
  \caption{(a) A schematic view of a typical tau decay (left) and hadronic jet (right).
  (b) Layer structure of a reconstructed $3\times3$ TOB matched to the visible component of a hadronically decaying tau lepton. 
The $x$ and $y$ axis are the \deta~and \dphi~directions and the vertical scale denotes the energy deposited per cell in GeV, also shown per electromagnetic (EM) and hadronic (Had) layers and for the total TOB. 
  (b) Transverse energy of reconstructed TOBs for signal and background samples. All reconstructed TOBs are shown for the background while only TOBs matched to the visible component of a hadronically decaying taus are shown for the signal.
  }
\end{figure}

To prepare the input dataset, we simulated the TOB reconstruction algorithm used in ATLAS with the granularity of the ATLAS Delphes datacard, which includes a simpler version of the ATLAS calorimeter, in which only a electromagnetic and an hadronic layers are present but with flexible granularity as the trigger towers used in the ATLAS L1Calo.
Thus, the TOB used in the following consists of a $d\times d$ structure of cells spanning through the two layers of the calorimeter, where $d$ takes values of 3, 5 and 9. The motivation for this approach is to evaluate the different architecture performances when increasing the complexity of data.
An example of a $2\times3\times3$ TOB is illustrated in Figure~\ref{fig:tob_visual}, where its total and per-layer energies are also shown. 

The signal dataset is specifically prepared to include TOBs that originate from hadronic decay of tau leptons. 
In order to achieve this, only the TOBs that are closest to the truth visible components of the tau lepton decays are taken into consideration, with a maximum of two TOBs per event. 
On the other hand, the background dataset consists of all TOBs that are reconstructed in the calorimeter, as a single TOB passing the selection criteria will activate the trigger and be recorded. 
In this work, the complete dataset comprises two types of events: background and signal.  Each `event' consists of a batch of TOBs, where the batch size may vary between events.
Figure~\ref{fig:pt_tob_comparison} depicts the energy distributions for both signal and background TOBs.

\section*{Supervised learning training}
\label{sec:dataPre}

The classification task addressed in this work belongs to the supervised learning (SL) category. The model is trained to learn a function that maps inputs to outputs using a labeled training dataset.
In the following, a single TOB is referred to as a `sample', representing the individual input provided to each model.
The problem is approached as a binary classification scenario with a traditional loss function - the Cross-Entropy (CE) metric. To optimize the CE loss function, we employ the (iterative) Stochastic Gradient Descent (SGD) algorithm across all the tested models. In terms of data balance, the ratio of signal to background TOBs was maintained consistently throughout our discussions, with approximately 70k signal TOBs and 200k background TOBs.

We explored three different learning methods: one is a classic machine learning (ML) model, namely a decision tree, while the other two are NN models inspired by the computer vision field - the field that deals with gaining high-level understanding from digital images or videos. 
Note that while the model inputs are quite similar to those of `standard' NN tasks, in our case there are strict latency and computational constraints that arise from the very high rate of events of the ATLAS experiment. Therefore, the goal is to build architectures as simple as possible that can still be implemented with FPGAs.

The first concept that we have explored is the Extreme Gradient Boosting (XGBoost) algorithm ~\cite{Chen_2016} which became well known in the ML community after being part of the winning solution of the Higgs Machine Learning Challenge in Kaggle~\cite{pmlr-v42-cowa14}. XGBoost is a gradient boosting tree model that usually involves a small number of parameters compared to NN models since its inputs are only high level features produced from the raw data, hence allowing a simpler FPGA implementation with minimal computational requirements and low latency inference time~\cite{Hong:2021snb}. The high level features we used for this model describe the most important information about the energy deposits and their positions in the TOB in a single vector of size 17. The strongest features according to feature importance methods were the total energy deposited in the TOB, total energy in each layer, average penetration depth of the energy deposits along the calorimeter, ratio of the cell energies squared to their respective volumes, first and second maximal energy deposits of the second layer and few ratio combinations of all the above-mentioned.

The second architecture is a multi-layer perceptron (MLP) NN, which has the ability to learn linear and non-linear relationships between data elements due to to the neuron’s activation function. Due to simplicity reasons, we used for the MLP a minimal structure of two hidden layers with (5,4) units each. Unlike in XGBoost, the NN employs the entire raw data elements in a flattened shape and not as the raw data structure tensor. In this kind of model, all energy evidences are transferred as inputs but the model does not get explicit spatial information regarding the energy deposit positions in the TOB.

The third approach we study relies on the shape of each TOB and therefore treats it as a tensor - similar to a dual channel matrix, where the pixel values are the energy deposits of the cells.     
For this `image like' perspective we built an architecture based on the  Residual Network (ResNet)~\cite{he2015deep}, which consists of skip-connections and contains nonlinearities (using the ReLU activation function as in MLP) and batch normalization layers in between. As in NN, the purpose of this architecture is to transfer energy evidence; however, the physical positions of each cell are also taken into account. 



\section*{Results}
\label{sec:results}

\begin{figure}[ht] 
    \centering
    \includegraphics[width=0.3\linewidth]{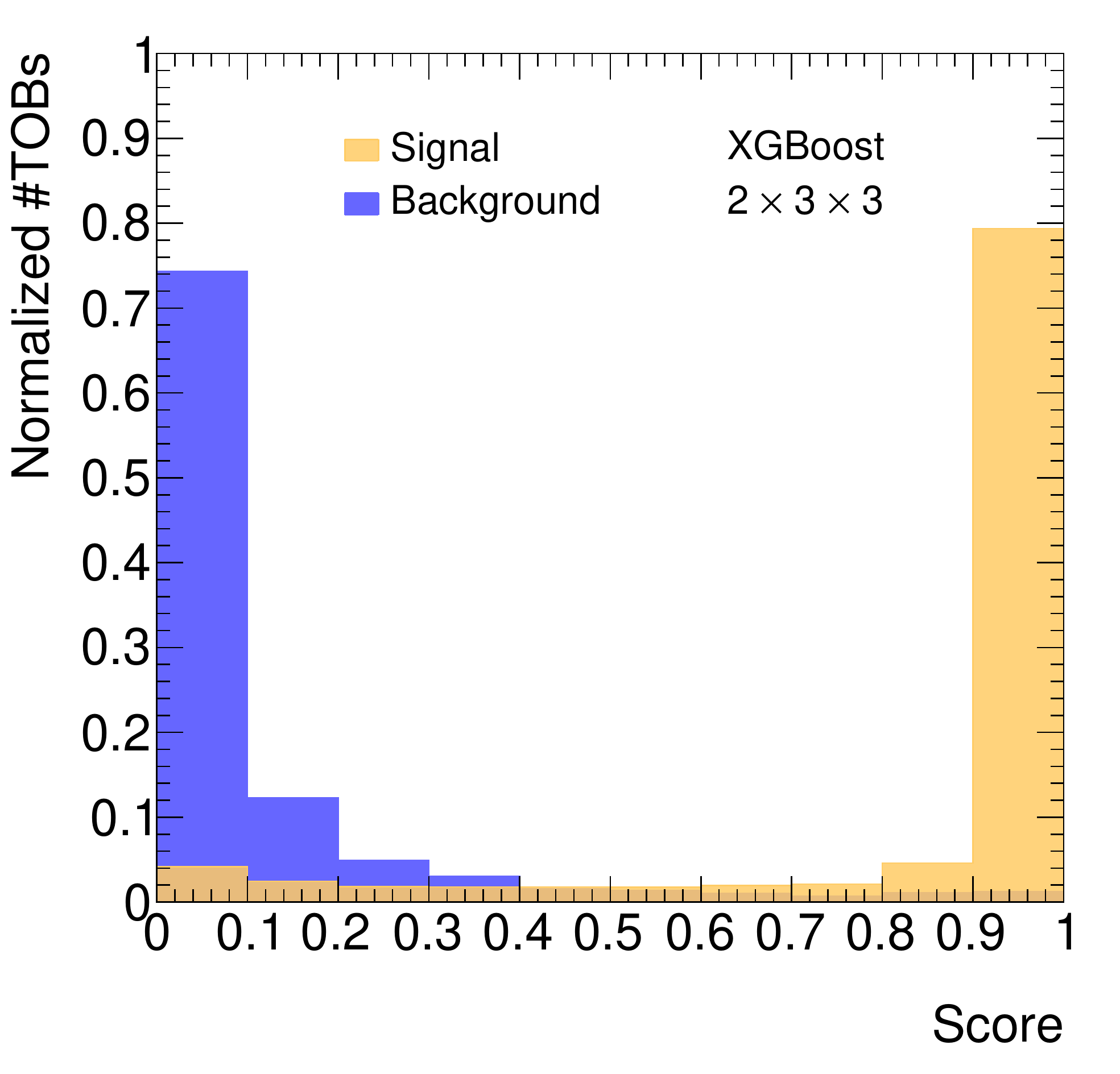}\includegraphics[width=0.3\linewidth]{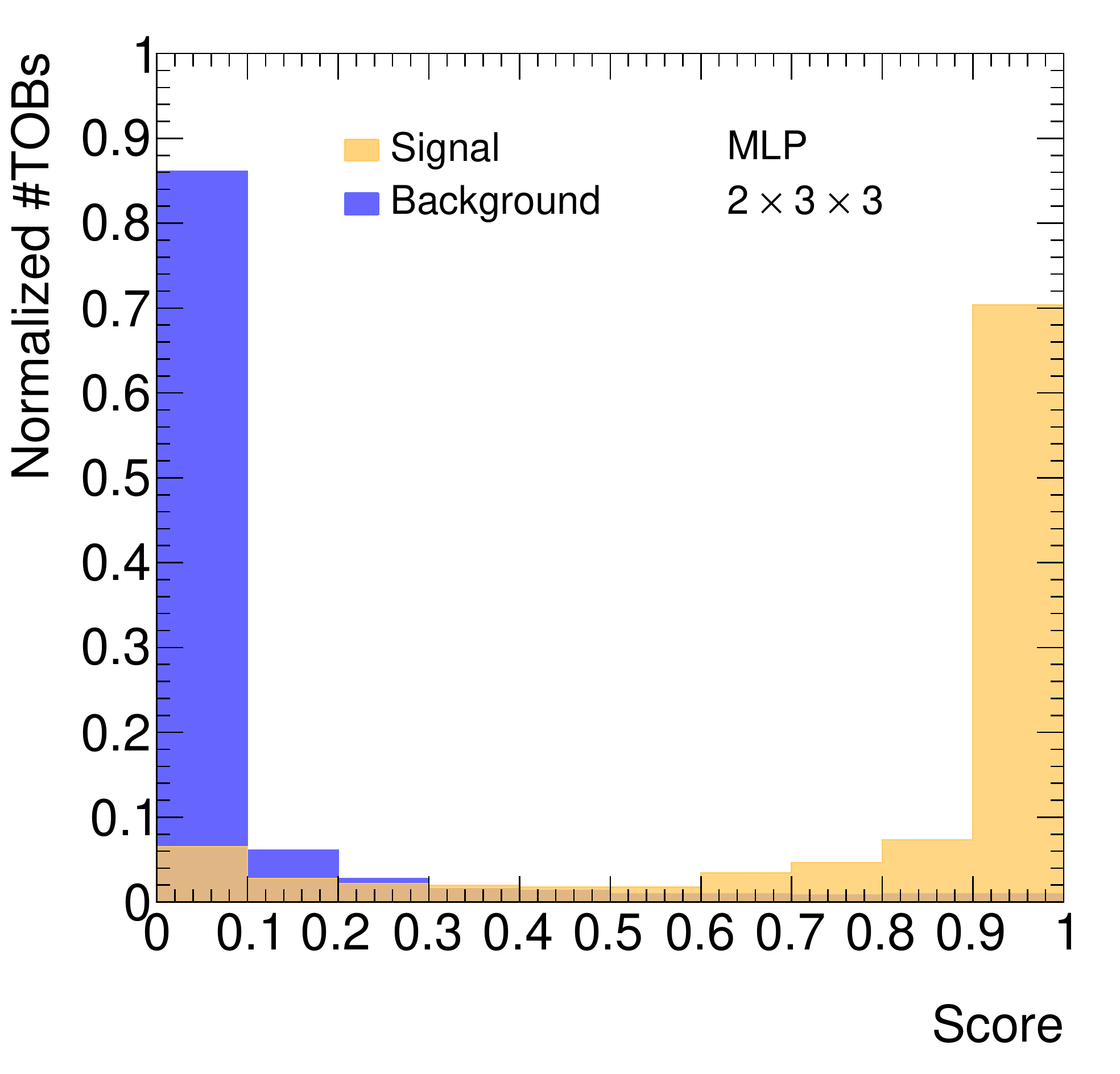}\includegraphics[width=0.3\linewidth]{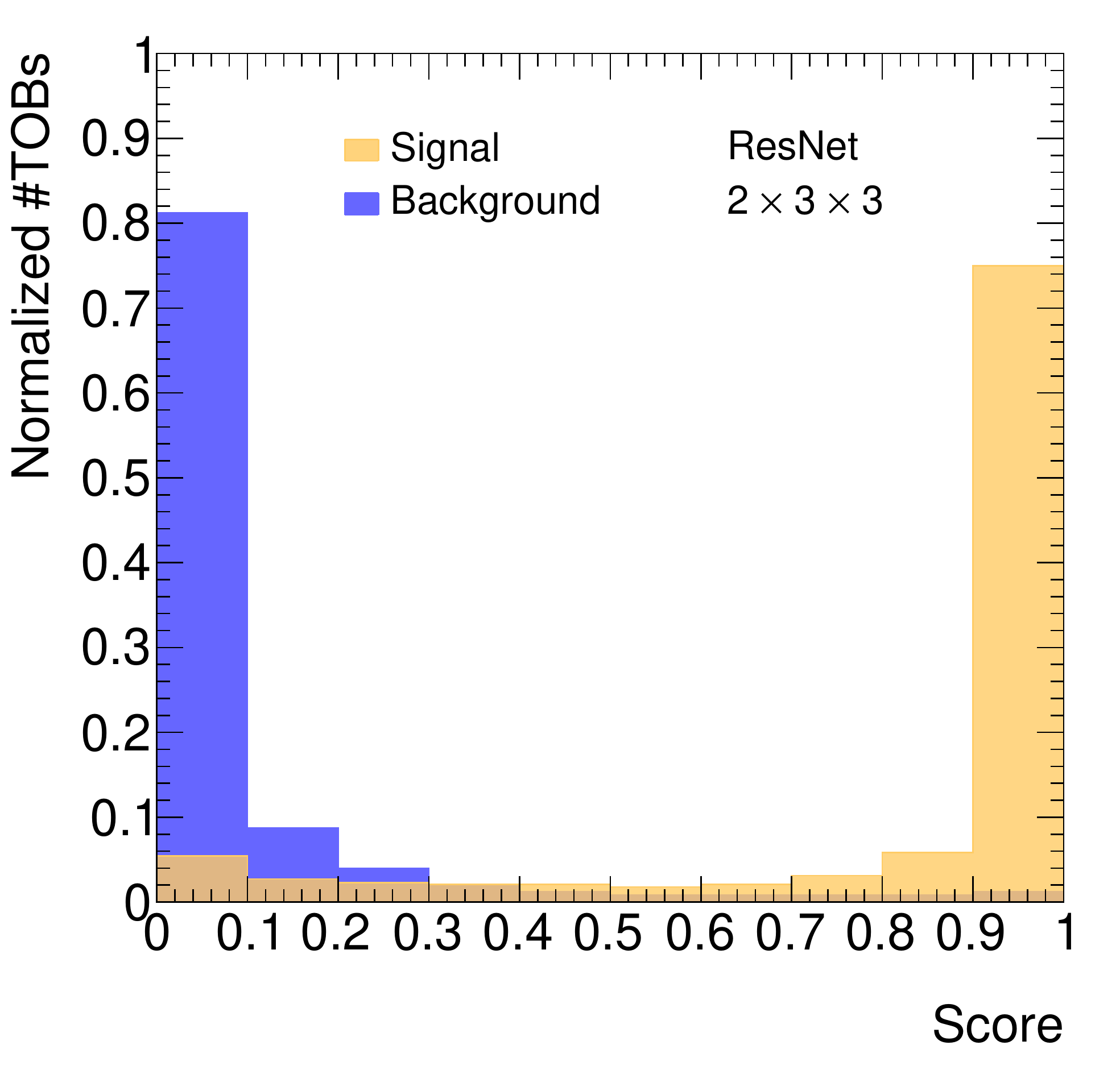}\\
        \centering
    \includegraphics[width=0.3\linewidth]{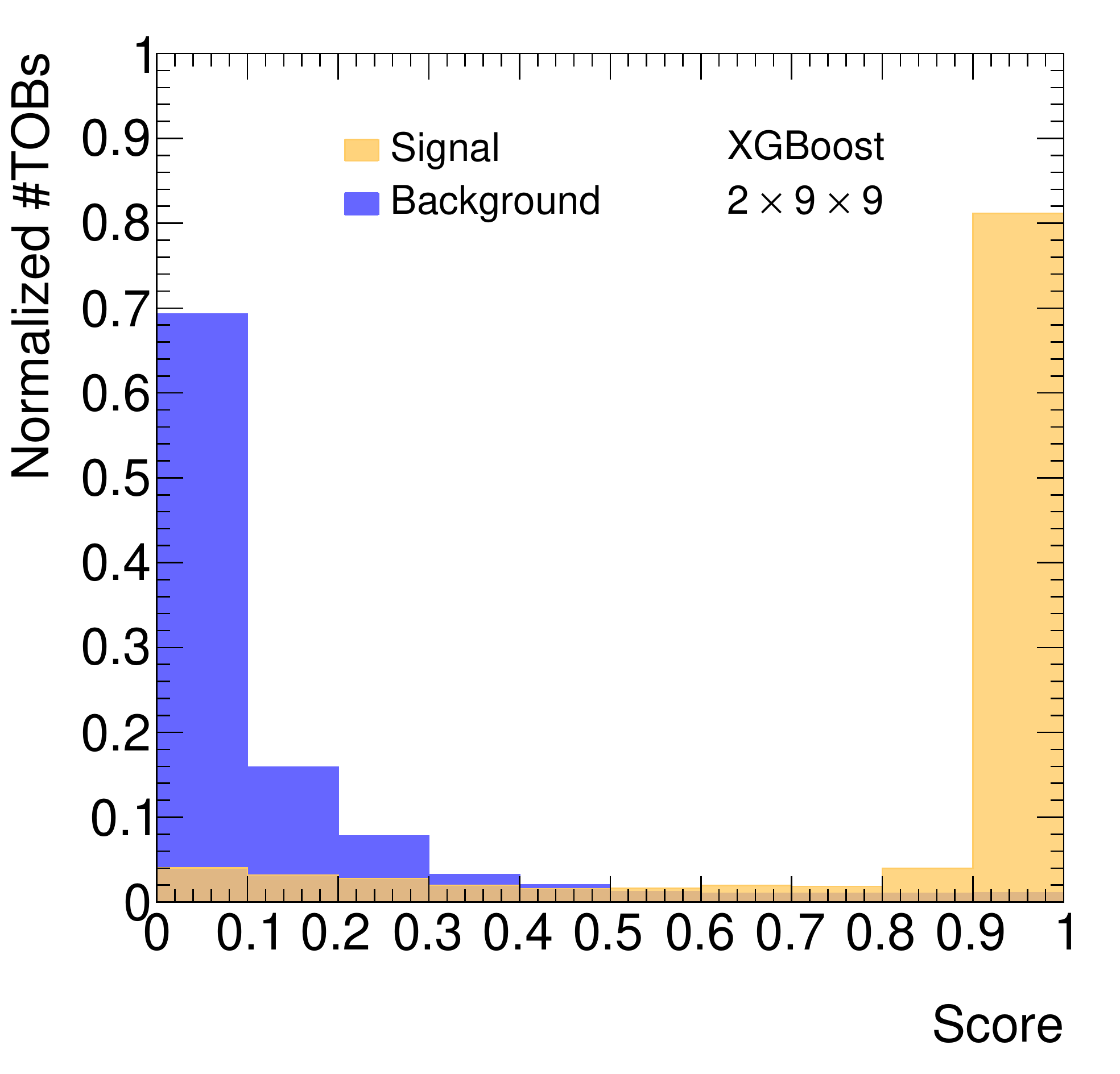}\includegraphics[width=0.3\linewidth]{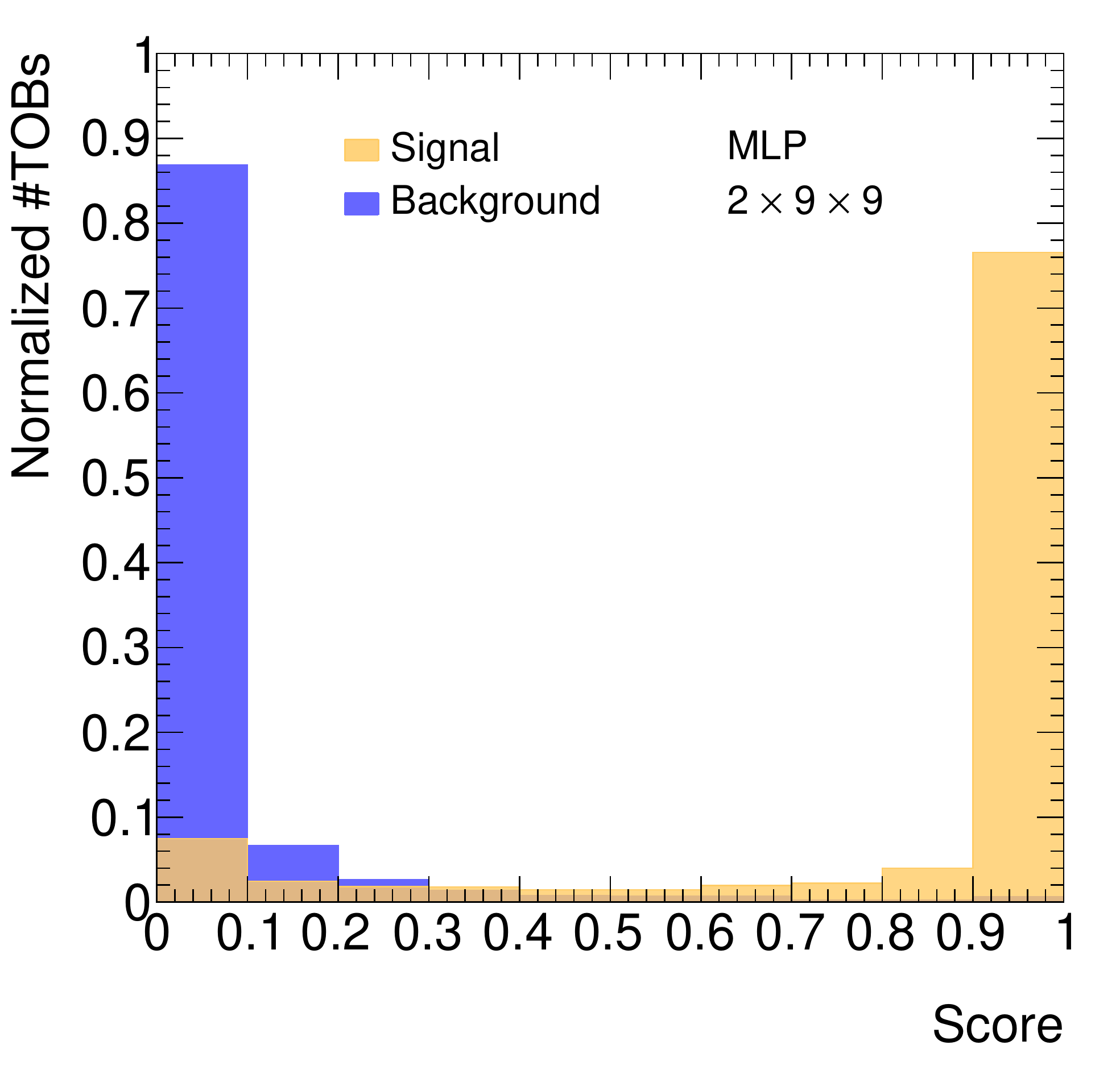}\includegraphics[width=0.3\linewidth]{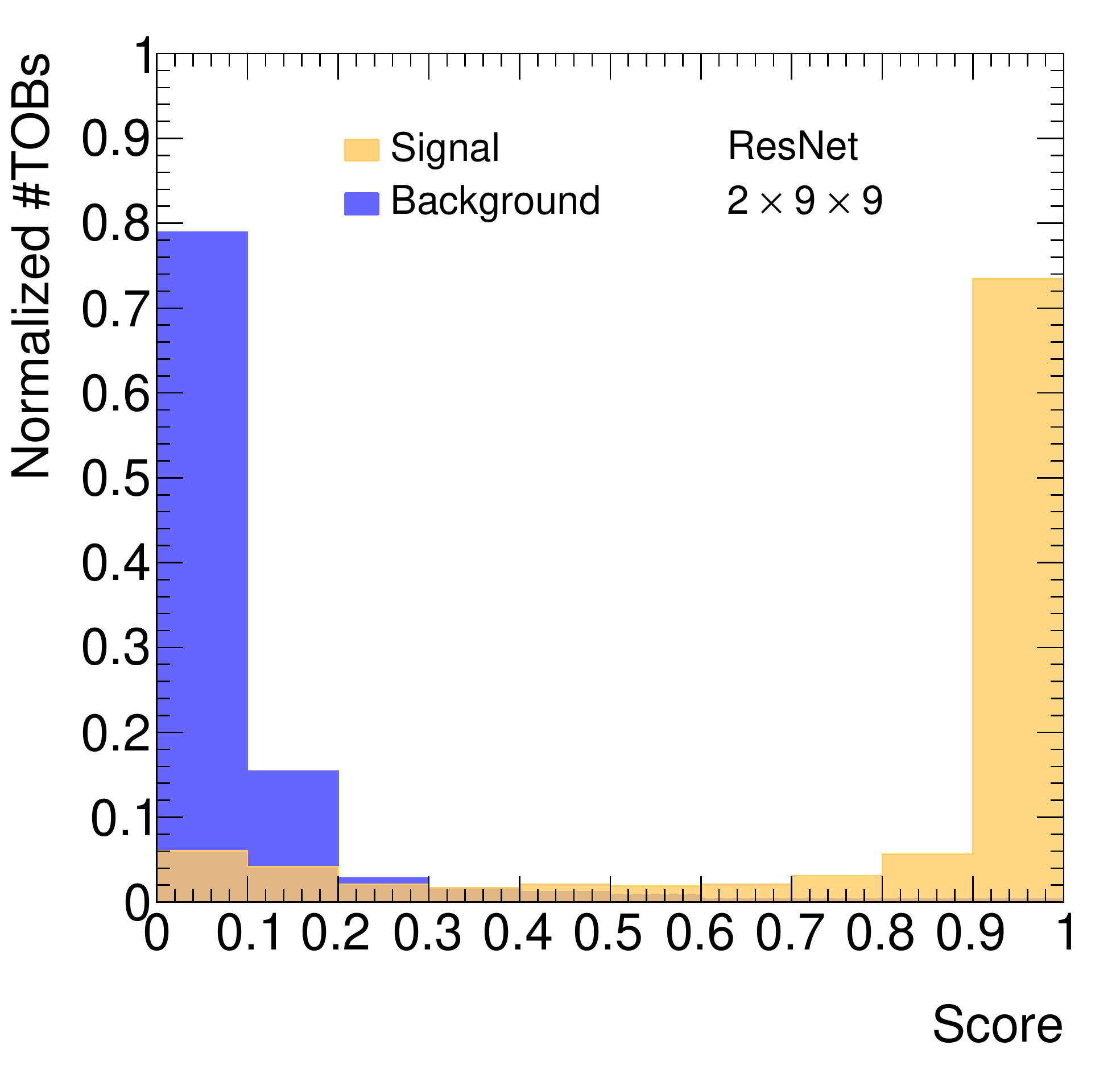}\\
    \caption{Trigger Objects (TOBs): Score distributions for signal (orange) and background (blue)  with  $2\times3\times3$ (top) and $2\times9\times9$ (bottom) dimensions for: left: XGBoost; center:  MLP; and right: ResNet;}
    \label{fig:results1}
\end{figure}

In this section we discuss the results for the aforementioned architectures using the three aforementioned datasets. All datasets include 200k background and around 70k signal samples (TOBs) but with one distinction - the dataset TOBs have a different cardinality in Hadronic and electo-magnetic layers so we get three tensor dimensions (this is 2 layers (EM,HAD), then the size of the towers): $2\times3\times3$, $2\times5\times5$ and $2\times9\times9$.
Prior to the evaluation of each classifier's performance, we first present the score distribution of the models. In Figure~\ref{fig:results1} we demonstrate that there is a good separation between signal and background distributions for all the architectures. However, there is a significant difference between XGBoost and NNs in the variance of the distributions for the high dimensional data structure; the XGB background score variance is higher than NNs' variance, a feature that might suggest a worse signal/background separation and therefore a lower classification quality for this kind of data. 

The performance is evaluated using multiple metrics; first we use the most common classification evaluation metrics, namely, the precision recall curve (PR) and the receiver operating characteristic curve (ROC), which show the performance of a classification model at all classification thresholds by presenting the precision against recall and true positive rate (TPR) against the false positive rate (FPR) respectively for each. For both curves we also measure the area under curve (AUC), which is an aggregated measurement of the prediction quality across all possible thresholds, with higher values corresponding to better classifiers. 

In addition to the previous metrics, we tested a more practical and unique metric, which is more adjusted to the needs of the hadron collider experiments such as ATLAS experiment: the Turn-on Curve (TOC), also known as an efficiency plot, and its AUC, the TOC-AUC. The TOC shows the efficiency as a function of a given observable obtained from a given model at the \textbf{event} level, as opposed to both the ROC and PR curves that quantify it at the TOB level. An event, composed of many TOBs, is classified as signal if even a single one of its TOBs is classified by the model as signal. Otherwise, it is classified as background. First, a model threshold is determined by the `fake rate' - the maximum number of background events that can be allowed to be mistakenly classified as signal. The fake rate is dictated by the maximum trigger bandwidth, i.e., the maximal rate of events that can be recorded. Then, given this model threshold, the model efficiency for the signal can be plotted separately for each \pT~range. Since it becomes easier to classify signal events as the $\tau$ energy increases, the TOC efficiency is expected to begin at 0 and approach 1. The quicker the `turn-on' response the more successful the model. In our case, the improvement can be seen in the low \pT~ranges of 10-20 GeV. This means using this method we can detect more low-\pT~taus that would otherwise be missed.

\begin{figure}[!ht]
    \centering
    \includegraphics[width=0.42\textwidth]{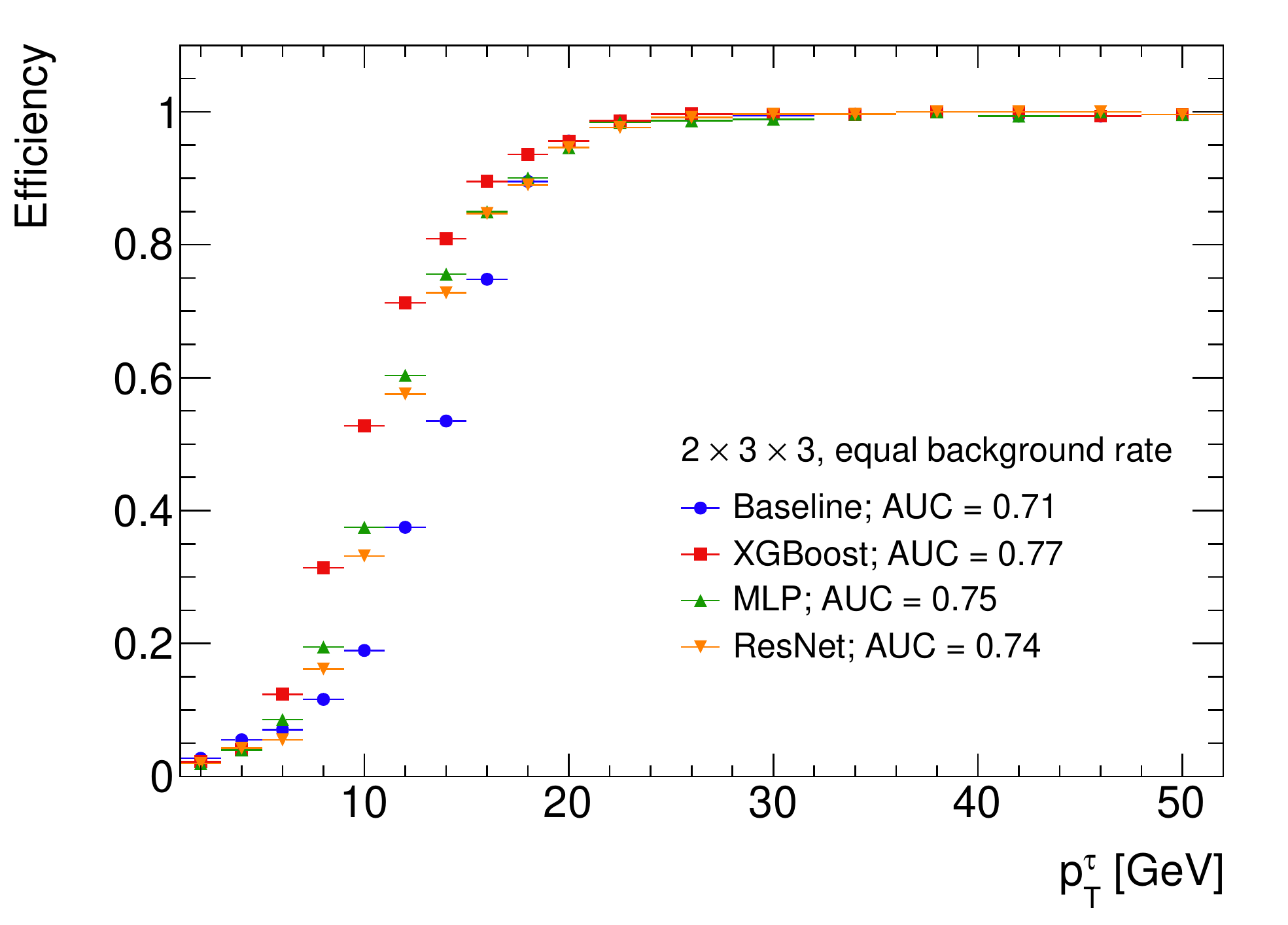}
    \includegraphics[width=0.42\textwidth]{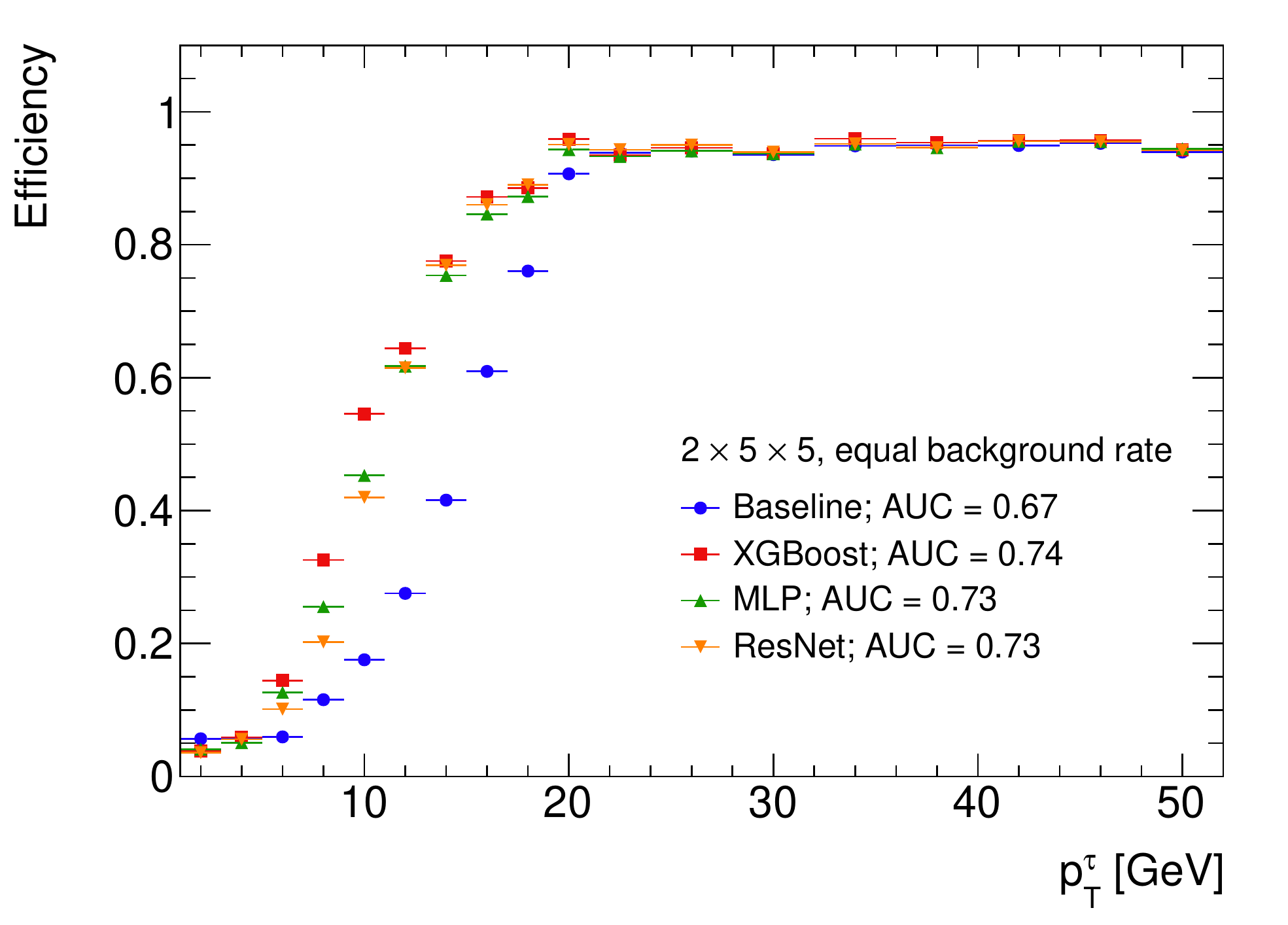}\\
    \includegraphics[width=0.42\textwidth]{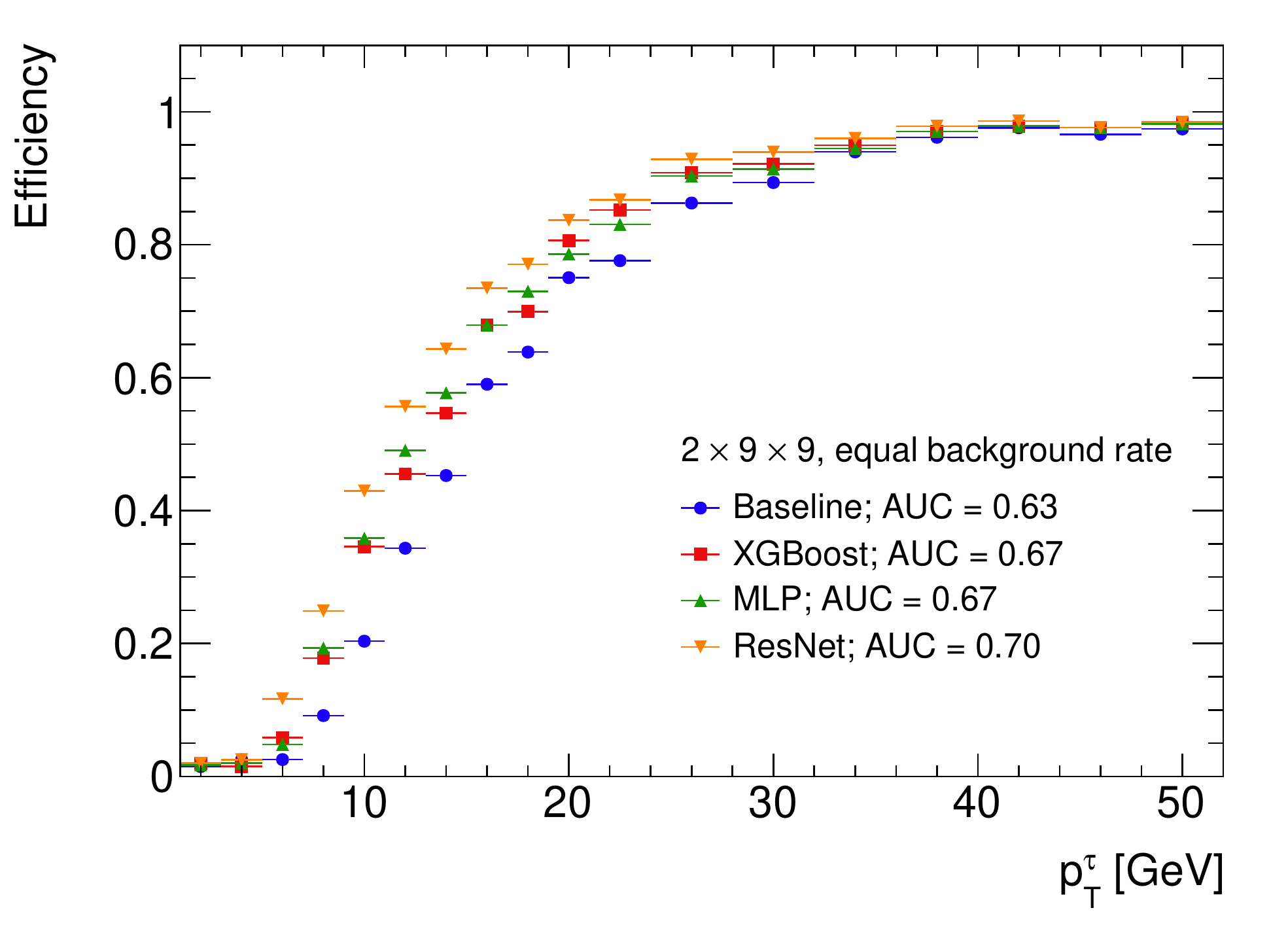}
    \caption{
    Turn on curve (TOC)  of the various architectures with different data dimension.}
    \label{fig:toc_curves}
\end{figure}

As can be seen in Table \ref{tab:metric_table}, the regular classification metrics reveal only tiny differences in performance between the three architectures in each of the data structures. It does not describe the performance vs \pT~and is also affected by the high imbalance of our dataset. This is in contrast to Figure \ref{fig:toc_curves} that presents the TOC curves, where we can evaluate the efficiency of every method: note that the baseline of our task is a classic algorithm and not an ML one; this algorithm simply looks for clusters in the TOB layers and above a certain threshold classifies the event as a signal. As we can see from the TOC, all ML algorithms have a much higher efficiency for \pT~below 20 GeV and are equal to the baseline performance above this range. The best algorithm in the low \pT~range is changing depending on the data structure we study: in the lowest TOB cardinality $2\times3\times3$, the XGBoost performance (in TOC terms for \pT~below 20 GeV) surpasses the NN models. For a slightly higher cardinality of $2\times5\times5$, XGBoost still takes the lead, however, the performance gap is reduced. Finally, when we run all models over the biggest data structure $2\times9\times9$, the algorithm ranking changes completely with ResNet performance at the top and XGBoost moving to the bottom. 

\begin{table}[h]
\centering
\caption{
ROC-AUC, PR-AUC and F1-MAX metrics and the number of parameters for each of the architectures and per every data dimension are shown. The metrics shown in bold denote which is the best architecture for each of the data dimensions.}
\begin{tabular}{|c | c|c|c|c|}
\hline
Metric & Granularity & XGBoost & MLP & ResNet \\ \hline
 &3$\times$3 & \textbf{0.969} & 0.967 & 0.968 \\ \cline{2-5} 
 \multirow{1}{*}{ROC\_AUC}& 5$\times$5&  \textbf{0.936} & 0.93 & 0.931  \\ \cline{2-5} 
 & 9$\times$9 & \textbf{0.967} & 0.963 & 0.964 \\ \hline
 
 & 3$\times$3& \textbf{0.956} & 0.952 & 0.952\\ \cline{2-5} 
\multirow{1}{*}{PR\_AUC} & 5$\times$5& \textbf{0.912} & 0.903 & 0.904 \\ \cline{2-5} 
&9$\times$9 & \textbf{0.955} & 0.95 & 0.952 \\ \hline

 & 3$\times$3 & \textbf{0.898} & 0.893 & 0.893  \\ \cline{2-5} 
\multirow{1}{*}{F1\_MAX}& 5$\times$5  &  \textbf{0.846} & 0.836 & 0.838 \\ \cline{2-5} 
& 9$\times$9  & \textbf{0.904} & 0.9 & 0.902 \\ \hline

&3$\times$3 & 1400 & 97 & 3778 \\ \cline{2-5} 
 \multirow{1}{*}{\# parameters}& 5$\times$5&  1700 & 225 & 10466  \\ \cline{2-5} 
 & 9$\times$9 & 1900 & 673 & 38050 \\ \hline

\end{tabular}
\label{tab:metric_table}
\end{table}

To provide some insight into the anticipated memory consumption of each architecture, we examine the count of trainable parameters. As illustrated in Table, the size of XGBoost  doesn't exhibit a significant increase with varying granularity sizes, as the number of input features remains constant across granularities. However, this trend doesn't apply to the Deep Neural Network (DNN). For instance, the MLP size experiences an impact primarily in the initial layer, which scales quadratically with the granularity dimension and is then multiplied by the number of TOB layers ($L \times d \times d $). In the case of ResNet, the influence of granularity dimension is even more pronounced, affecting not only the first layer but also subsequent layers.
While there remain numerous unexplored parameters in determining the optimal algorithm for a given hardware, XGBoost and MLP algorithms present themselves as promising baselines. These algorithms offer the advantage of adjustability based on specific hardware constraints. Subsequently, through a meticulous fine-tuning process, it becomes feasible to evaluate whether the heightened complexity of DNNs is warranted from an efficiency perspective. This evaluation can also shed light on the feasibility of implementing such intricate models for exceedingly high granularity scenarios.

In summary of our findings, we assessed two types of metrics: the conventional binary classification metrics including ROC\_AUC, PR\_AUC, and F1\_MAX, and the TOC\_AUC metric, which relies on \pT. While binary classification metrics excel in straightforward classification evaluation, the TOC\_AUC metric stands out in tasks involving low \pT regime classification. Regarding binary classification metrics, both XGBoost and NN demonstrate comparable performance, with slight variations based on grid size. However, when considering the TOC\_AUC metric, it proves notably advantageous for NN applications in larger grid sizes. Interestingly, a trend emerges within this metric: as the grid size increases, the performance of ResNet experiences significant enhancement. This suggests that in scenarios where the TOC\_AUC metric is crucial, XGBoost is preferable for smaller grid sizes, while NN architectures such as ResNet become increasingly favorable as the grid size expands.

\section*{Conclusions}
\label{sec:conclusion}

The growing rate of $pp$ events at the LHC is vital for searching for new phenomena. However, the increasing pace  of collected data poses challenges in filtering the relevant events.
A novel $\tau$ triggering technique is introduced that is able to significantly improve the ability to trigger on low \pT~hadronically decaying $\tau$s.
We introduce a decision tree trained with XGBoost and advanced deep learning techniques such as MLP and ResNet, and demonstrate their ability at different complexity levels of data structures.
In the future, LHC experiments can utilize these techniques to accommodate the anticipated conditions, which will involve higher detector granularity and more intricate data structures.
For most of the energy range considered, ResNet is found to be the best performing technique for high dimensional structure while for low complexity data, a classic ML approach like XGBoost gives the best performance. We believe that these observations are relevant also to other scientific problems beyond tau triggers and can guide other researchers in the selection of the machine learning tool to be used for their data.

\bibliography{nntau}

\section*{Acknowledgements}

We thank Ben Carlson and Stephen T. Roche for discussions and guidance on how to handle the Delphes samples, as well as providing us with comments on the draft.
This project has received funding from the European Research Council (ERC) under the 
European Union’s Horizon 2020 research and innovation programme (Grant agreement No. 945878)

\section*{Author contributions statement}

M.Y. analysed the data, paper editor, U.B. analysed the data L.P.D. samples prepropcessing, paper editor, B.C. samples preprocessing. R.G. a machine learning exeprt, with L.B and E.E.,  experimental particle physicists, jointly initiated and supervised the project. All authors reviewed the manuscript.

\section*{Additional information}
 \textbf{Accession codes} The complete implementation of this work can be accessed at: \href{https://github.com/MaayanLucyYaari/tua-trigger.git}{https://github.com/MaayanLucyYaari/tua-trigger.git};
\textbf{Competing interests} The authors declare no competing interests..

\end{document}